\documentstyle[psfig,12pt,citesort]{article}

\def \ov {\over}

\newcounter{subequation}[equation]

\def\e{\epsilon}

\newcommand{\be}{\begin{equation}}  
\newcommand{\ee}{\end{equation}}  
\newcommand{\eel}[1]{\label{#1}\end{equation}}  
\newcommand{\bea}{\begin{eqnarray}}  
\newcommand{\eea}{\end{eqnarray}}   
\newcommand{\eeal}[1]{\label{#1}\end{eqnarray}}  
\makeatletter  
\def\thesubequation{\theequation\@alph\c@subequation}  
\def\@subeqnnum{{\rm (\thesubequation)}}  
\def\slabel#1{\@bsphack\if@filesw {\let\thepage\relax  
   \xdef\@gtempa{\write\@auxout{\string  
      \newlabel{#1}{{\thesubequation}{\thepage}}}}}\@gtempa  
   \if@nobreak \ifvmode\nobreak\fi\fi\fi\@esphack}  
\def\subeqnarray{\stepcounter{equation}  
\let\@currentlabel=\theequation\global\c@subequation\@ne  
\global\@eqnswtrue \global\@eqcnt\z@\tabskip\@centering\let\\=\@subeqncr  
  
$$\halign to \displaywidth\bgroup\@eqnsel\hskip\@centering  
  $\displaystyle\tabskip\z@{##}$&\global\@eqcnt\@ne  
  \hskip 2\arraycolsep \hfil${##}$\hfil  
  &\global\@eqcnt\tw@ \hskip 2\arraycolsep  
  $\displaystyle\tabskip\z@{##}$\hfil  
   \tabskip\@centering&\llap{##}\tabskip\z@\cr}  
\def\endsubeqnarray{\@@subeqncr\egroup  
                     $$\global\@ignoretrue}  
\def\@subeqncr{{\ifnum0=`}\fi\@ifstar{\global\@eqpen\@M  
    \@ysubeqncr}{\global\@eqpen\interdisplaylinepenalty \@ysubeqncr}}  
\def\@ysubeqncr{\@ifnextchar [{\@xsubeqncr}{\@xsubeqncr[\z@]}}  
\def\@xsubeqncr[#1]{\ifnum0=`{\fi}\@@subeqncr  
   \noalign{\penalty\@eqpen\vskip\jot\vskip #1\relax}}  
\def\@@subeqncr{\let\@tempa\relax  
    \ifcase\@eqcnt \def\@tempa{& & &}\or \def\@tempa{& &}  
      \else \def\@tempa{&}\fi  
     \@tempa \if@eqnsw\@subeqnnum\refstepcounter{subequation}\fi  
     \global\@eqnswtrue\global\@eqcnt\z@\cr}  
\let\@ssubeqncr=\@subeqncr  
\@namedef{subeqnarray*}{\def\@subeqncr{\nonumber\@ssubeqncr}\subeqnarray}  
  
\@namedef{endsubeqnarray*}{\global\advance\c@equation\m@ne%  
                           \nonumber\endsubeqnarray}  
  
\makeatletter \@addtoreset{equation}{section} \makeatother  
\renewcommand{\theequation}{\thesection.\arabic{equation}}  
\catcode`\@=11  
  
\newcount\hour  
\newcount\minute  
\newtoks\amorpm \hour=\time\divide\hour by 60\minute  
=\time{\multiply\hour by 60 \global\advance\minute by-\hour}  
\edef\standardtime{{\ifnum\hour<12 \global\amorpm={am}%  
        \else\global\amorpm={pm}\advance\hour by-12 \fi  
        \ifnum\hour=0 \hour=12 \fi  
        \number\hour:\ifnum\minute<10  
        0\fi\number\minute\the\amorpm}}  
\edef\militarytime{\number\hour:\ifnum\minute<10 0\fi\number\minute}  
  
\def\draftlabel#1{{\@bsphack\if@filesw {\let\thepage\relax  
   \xdef\@gtempa{\write\@auxout{\string  
      \newlabel{#1}{{\@currentlabel}{\thepage}}}}}\@gtempa  
   \if@nobreak \ifvmode\nobreak\fi\fi\fi\@esphack}  
        \gdef\@eqnlabel{#1}}  
\def\@eqnlabel{}  
\def\@vacuum{}  
\def\marginnote#1{}  
\def\draftmarginnote#1{\marginpar{\raggedright\scriptsize\tt#1}}  
\def\draft{  
        \pagestyle{plain}  
        \overfullrule=2pt  
        \oddsidemargin -.5truein  
        \def\@oddhead{\sl \phantom{\today\quad\militarytime} \hfil  
        \smash{\Large\sl DRAFT} \hfil \today\quad\militarytime}  
        \let\@evenhead\@oddhead  
        \let\label=\draftlabel  
        \let\marginnote=\draftmarginnote  
        \def\ps@empty{\let\@mkboth\@gobbletwo  
        \def\@oddfoot{\hfil \smash{\Large\sl DRAFT} \hfil}  
        \let\@evenfoot\@oddhead}  
  
\def\@eqnnum{(\theequation)\rlap{\kern\marginparsep\tt\@eqnlabel}%  
        \global\let\@eqnlabel\@vacuum}  }

\renewcommand{\theequation}{\thesection.\arabic{equation}}  
\renewcommand{\thefootnote}{\fnsymbol{footnote}}

\def\appendix#1{  
  \addtocounter{section}{-3}  
  \setcounter{equation}{0}  
  \renewcommand{\thesection}{\Alph{section}}  
  \section*{Appendix \thesection\protect\indent \parbox[t]{11.15cm}  
  {#1} }  
  \addcontentsline{toc}{section}{Appendix \thesection\ \ \ #1}  
  }  
\def \ov {\over}

\def\e{\epsilon}

\def\e{\epsilon}  
\def\m{\mu}  
  
\def\a{\alpha}  
\def\b{\beta}

\def\be{\begin{equation}}  
\def\ee{\end{equation}}

\def \m {\mu}

\date{}  
\begin{document}  
%\draft  
  
\begin{titlepage}

\hfill hep-th/0208066  
  
\hfill MCTP-02-41\\  
  
\hfill PUPT-2046\\  
\begin{center}  
%\vskip 2.5 cm \vskip 1 cm  

{\Large \bf  Strings in RR Plane Wave  Background}  
  
\vskip .3cm  
  
{\Large \bf  at Finite Temperature}

\vskip .7 cm

\vskip 1 cm  
  
{\large   Leopoldo A. Pando Zayas${}^{1}$ and Diana Vaman${}^{2}$}\\  
  
\end{center}  
  
\vskip .4cm \centerline{\it ${}^1$ Michigan Center for Theoretical  
Physics}  
\centerline{ \it Randall Laboratory of Physics, The University of  
Michigan}  
\centerline{\it Ann Arbor, MI 48109-1120}

\vskip .4cm \centerline{\it ${}^1$ School of Natural Sciences}  
\centerline{ \it Institute for Advanced Study}  
\centerline{\it Princeton, NJ 08540}

\vskip .4cm \centerline{\it ${}^2$ Department of Physics}  
\centerline{ \it Princeton University}  
\centerline{ \it Princeton, NJ 08544}

\vskip 1 cm

\vskip 1.5 cm  
  
\begin{abstract}  
We calculate the thermal partition function in the   
canonical ensemble for type IIB superstrings  
in the plane wave background with constant null R-R 5-form.   
The Hagedorn temperature is found to   
be higher than the corresponding value for  strings in  
flat space. In the limit corresponding to the weakly coupled   
field theory we find that   
the Hagedorn temperature is pushed to infinity.  
The key property of strings in the plane wave background  
under investigation  
on which our result relies is that the effective mass of the bosonic  
and fermionic coordinates in the light-cone gauge is proportional to the  
momentum $p^+$.  The free energy    
is finite as the Hagedorn temperature is approached from below,
suggesting a possible phase transition.

\end{abstract}  
\vskip 1cm  
\mbox{~~~~~~~email addresses: lpandoz@ias.edu, ~  
dvaman@feynman.princeton.edu}  
\end{titlepage}  
\setcounter{page}{1} \renewcommand{\thefootnote}{\arabic{footnote}}  
\setcounter{footnote}{0}  
  
\def \N{{\cal N}} \def \ov {\over}  
  
\setcounter{page}{1} \renewcommand{\thefootnote}{\arabic{footnote}}  
\setcounter{footnote}{0}  
  
\def \N{{\cal N}} \def \ov {\over}

%%%%%%%%%%%%%%%%%%%%%%%%%%%%%%%%%%%%%%%%%%%%%%%%%%%%%%%%%%%%%%%%%%%%%  
\section{Introduction and Summary}  
%%%%%%%%%%%%%%%%%%%%%%%%%%%%%%%%%%%%%%%%%%%%%%%%%%%%%%%%%%%%%%%%%%%   

The AdS/CFT correspondence has given a concrete realization of the   
connection between gauge theories and string theory   
in the context of ${\cal N}=4$ supersymmetric Yang-Mills   
and type IIB string theory on $AdS_5\times S^5$ \cite{ads}. One of   
the immediate implications of the   
correspondence, pointed out by Witten  in \cite{wittenads},   
is the relation between the corresponding theories at nonzero temperature.   
 Namely, it was conjectured  
that a direct relation exists between a nonzero temperature field theory  
and the properly thermalized supergravity background. This  
correspondence was effectively anticipated in the series of papers  
\cite{igort}. A remarkable feature of the nonzero temperature correspondence  
is that it involves nonsupersymmetric theories on both sides and it  
is therefore a more dynamical connection than the usual zero-temperature   
correspondence.  
  
Recently, Berenstein, Maldacena and Nastase (BMN)  \cite{bmn}  
put a new twist on the  
AdS/CFT correspondence and     
proposed a gauge theory interpretation of the Penrose-G\"uven limit  
of $AdS_5\times S^5$ \cite{blau}. The IIB maximally supersymmetric   
plane wave background with constant null RR five form   
resulting from the limit is   
\bea  
ds^2&=&-2dx^+dx^- - \mu^2 x_ix^i (dx^+)^2+ dx_idx^i, \nonumber \\  
F_{+1234}&=&F_{+5678}\,\,\,=\,\,\,2\mu.  
\eea  
In this limit  a particular sector of  ${\cal N}=4$ SYM survives in the  
gauge theory side while the string theory is solvable    
in the light-cone   
gauge \cite{metsaev,metse}.   
One particularly interesting observation that follows from the study  
of BMN operators is that certain quantities, such as the dimensions of  
the BMN operators, when calculated using perturbative gauge theory   
in the effective 't Hooft coupling $(\lambda')$   
match precisely the string theory prediction which is reliable at   
strong coupling \cite{gauge}. This provides an example of smooth  
interpolation between the weak and strong coupling regimes of  
a sector of ${\cal N}=4$ SYM. Additional exploration of more   
complicated observables was presented in \cite{igor}.

In the context of the Penrose limit of $AdS_5\times  
S^5$, it is very natural to pose a similar question about the nonzero  
temperature correspondence. The present situation is very peculiar however.   
An important ingredient introduced in the analysis of \cite{wittenads} is   
the consideration of all backgrounds with the same asymptotic behavior. In the   
concrete case of ${\cal N}=4$ SYM this implies the need to consider  
at least thermalized $AdS_5$ and the Schwarzschild-AdS black hole.   
The inclusion of these two backgrounds implies   
the presence of Hawking-Page phase transitions.   
The question of the proper thermalization of the plane   
wave background is ambiguous. Moreover,   
the Penrose  
limit of the Schwarzschild-AdS black hole does not seem to naturally  
determine a horizon and therefore it lacks the notion of temperature   
\cite{cobi}. Thus, the supergravity  
approach seems to be at least not straightforward. On the other hand,  
the fact that string theory on the plane wave can be quantized in the  
light-cone gauge gives us the possibility of exploring some of the  
corresponding thermodynamic properties.    
  
Strings in flat space exhibit a Hagedorn density of states  
$\rho(M)=M^{-a} \exp(bM)$, with $a$ and $b$ constants characterizing   
the various string models.  
Given the exponential growth in the density of states, the theory   
cannot be defined above a critical temperature $T_H\sim 1/(l_s b)$.     
This behavior has inspired a lot of speculations; one analogy that   
is frequently made is between the Hagedorn transition   
of strings and the confinement/deconfinement phase   
transition of QCD. It is natural to return to this  
question within the context of the BMN limit of  
${\cal N}=4$ SYM since it provides an example of gauge/string relation on  
one side of which we have an exactly solvable string theory.   
  
At first sight   
it seems very plausible that the high-temperature observables of string  
theory in the plane wave  
background have to be essentially the same as in flat space. An intuitive  
way to think about this is by using the fact that  
ultimately the main contribution to high temperature effects comes  
from the behavior of highly excited strings. Adding a mass term $(m)$   
to the bosons  
does not seem to be a strong enough modification since,  
for any $m$ we have that for large enough excitation levels ($n$),   
the expression  
for the frequencies   
$\sqrt{n^2 + m^2}$ will be better and better approximated  
by the flat space expression:   $n$. However, a  key property of  
strings in the plane-wave  background  
that invalidates the above reasoning  is that in the light-cone gauge the mass  
parameter is {\it not constant}, but $p^+$ dependent:  
$m=2\pi \alpha' \, \mu \, p^+$. We will show that this  
explicit dependence on $p^+$  
when properly taken into consideration modifies the modular  
properties of the  
partition function in a crucial way and in particular  
is responsible for the Hagedorn temperature being higher than the  
corresponding value for strings in flat space.  
  
To make the connection to the BMN sector more precise recall   
the following relations between the string and the gauge   
theory quantities $\m \, \alpha' p^+ = {J\over \sqrt{\lambda}},   
\,\, {2 p^-\over \m}= \Delta - J,$  
where $J$ is the R charge, $\Delta$ is the conformal dimension and   
$\lambda$ is the 't Hooft coupling.   The BMN sector is singled   
out  by considering operators with fixed $p^+$ and finite $p^-$.   
Having fixed $p^+$ means considering operators whose R charge J   
grows as $\sqrt{N}$ as $N\to \infty$. We,  
by integrating over $p^+$,    
are forced to effectively consider a larger sector than   
the BMN sector. This fact obscures the   
direct relevance of our string calculation for the gauge theory side.   
  
Leaving aside the usual difficulties associated with   
defining thermodynamics in gravitational backgrounds, in this paper   
we compute the   
thermal partition function of the IIB superstring  
in the plane wave background with constant null R-R 5-form.   
Moreover, we restrict ourselves to the leading genus one contribution  
to the thermal partition function. We pay special attention to the  
Hagedorn temperature which has now a parametric dependence on  
$\mu$. We find that for $\m\ne 0$   
the Hagedorn temperature is higher than the corresponding value in   
flat space  
\be  
-\frac{\beta_H^2}{2\pi\,\alpha'}-16 \pi\gamma_0(\mu\,\beta_H)  
+16 \pi \gamma_{1/2}  
(\mu\,\beta_H)=0,  
\ee  
where $\gamma_0$ and $\gamma_{1/2}$ are the zero-point energies   
of integer and half-integer moded massive oscillators respectively.   
In the limit $\m l_s \to 0$, the first correction to the flat space   
value of the Hagedorn temperature is $T_H\approx1/2\pi l_s \,+\, 4\mu$.  
In the ``weak field theory limit'' $(\mu l_s \to \infty)$ we   
find that the Hagedorn temperature is pushed to infinity. We also   
provide an  approximate expression for the free   
energy near the Hagedorn temperature assuming that the main contribution   
comes from the one-string sector. A saddle point 
evaluation shows that the free energy and  is finite as we approach the Hagedorn temperature from   
below. Therefore a phase transition could take place at $T_H=1/\beta_H$ .   
Given that in the canonical ensemble 
the energy fluctuations are notoriously large nearby $T_H$, 
the answer to the true nature (limiting  
or phase transition) of the Hagedorn temperature is hold by a 
complementary picture to the   
one presented here, namely  by the microcanonical ensemble.

The paper is organized as follows.   
Section 2 contains a short review of    
the canonical ensemble for strings in   
the light-cone gauge. Section 3 contains the essential   
ingredients in the calculation of the thermal partition   
function for the Green-Schwarz (GS) strings in the plane wave   
background.   
Throughout section 3 we make explicit how our results generalize   
those for GS strings in flat space.   
In section 4 we assemble all the ingredients   
and discuss the two natural limiting cases: flat space $(\m l_s \to 0)$ and   
the weak field theory $(\mu l_s \to \infty)$ and discuss the thermodynamic   
quantities of interest (free energy and specific heat).

%%%%%%%%%%%%%%%%%%%%%%%%%%%%%%%%%%%%%%%%%%%%%%%%%%%%%%%%%%%%%%%%%%%  
\section{Review of the canonical ensemble of strings in light-cone}  
%%%%%%%%%%%%%%%%%%%%%%%%%%%%%%%%%%%%%%%%%%%%%%%%%%%%%%%%%%%%%%%%%%%%  
  
Since in the following sections we need to address the thermal properties of  
strings whose action is known only in the  light-cone gauge-fixed form,  
and therefore a covariant treatment is not available, we begin with a brief  
review of the canonical ensemble of light-cone strings  
\cite{a1,a2}.  
As explained in \cite{a1}, the finite temperature one-string  
partition function is defined by  
\be  
Z_1(\beta)=\rm{Tr}\, e^{-\b\, P^0}= \rm{Tr}\, e^{-\beta (P^+ + P^-)},  
\ee  
where $\beta $ is the inverse temperature,  
$P^+ |\Phi>= p^+ |\Phi>$ is the light-cone momentum, and  
$P^- |\Phi>= \frac{1}{ p^+}\left(H_{lc}|\Phi>\right)$ with $H_{lc}$ being  
the light-cone Hamiltonian.  
Thus the partition function can be written as  
\be  
Z_1(\beta)=l_s\int dp^+ e^{-\beta p^+} z_{lc}(\beta/p_+).  
\ee  
For open strings, the partition function is evaluated over a cylindrical  
worldsheet, with periodic time $t\sim t+\beta/p^+$, while  
for closed strings we must additionally impose the level-matching constraint:  
equal momentum carried by the left and right oscillators.  
This is usually done by means of a Lagrange multiplier  
\be  
Z_1(\beta)=l_s\int dp^+ \int d\lambda e^{-\beta p^+}  
Tr e^{-\frac\beta{p^+}H_{lc} + 2\pi i\lambda (N_L-N_R)}.  
\ee  
From the point of view of the light-cone partition function, the effect of  
the angle $\lambda$ is to twist the ends of the cylinder  
before identifying them.   
  
Another useful way to approach the partition function is by way of  
the path integral formalism, with the light-cone partition function given  
by   
\be  
z_{lc}=\int {\cal D} X e^{-\int_T d^2 z L[X(z)]},  
\ee  
and where the worldsheet action is integrated    
over a torus $T$ with modular parameter  
\be  
\tau=\tau_1+i\tau_2=  
\lambda+i\frac{\beta}{2\pi p^+}.  
\ee  
One of the advantages of the path integral approach is that   
it makes the modular properties of the partition function manifest. We  
will repeatedly exploit this fact in the following sections.   
  
Having defined the one-string partition function it is natural to   
define the partition function for a gas   
of strings at constant temperature (canonical ensemble).   
The partition function for a gas of strings is built out of the   
one-string partition function   
through further exponentiation  
\be  
\ln Z(\beta)=\frac 12\sum_{r=1}^\infty [1-(-1)^r]\frac {Z_1(\beta r)}{r},  
\label{z}  
\ee  
for the supersymmetric case and   
$\ln Z(\beta)=\frac 12\sum_{r=1}^\infty {1\over r}Z_1(\beta r)$ for the   
bosonic case.   
The thermodynamical potential associated with the canonical ensemble   
is the free energy  
\bea  
F(\beta) &=&-{1\over \beta}\ln Z(\beta).  
\eea  
Another thermodynamical quantity of interest is the specific heat  
\bea  
c_V&=&\beta^2{\partial^2 \over \partial \beta^2} \ln Z(\beta).  
\eea  
At temperatures higher than the Hagedorn temperature $T_H$   
the free energy diverges.  
If the free energy remains finite while $T_H$ is approached from below  
there is the possibility of a phase transition. On the other hand,  
if the free energy and specific heat are infinite, the Hagedorn   
temperature is a limiting (maximal) temperature.

%%%%%%%%%%%%%%%%%%%%%%%%%%%%%%%%%%%%%%%%%%%%%%%%%%%%%%%%%%%%%%%%%%%  
\section{Partition function  of GS strings in the pp-wave background}  
%%%%%%%%%%%%%%%%%%%%%%%%%%%%%%%%%%%%%%%%%%%%%%%%%%%%%%%%%%%%%%%%%%%%%%  

The most straightforward way to obtain the partition function is   
through the Hamiltonian.   
The light-cone Hamiltonian of the first-quantized  
GS IIB string in the plane wave background with constant    
null R-R five-form background is given by \cite{metsaev,metse}:  
  
\bea  
H_{lc}&=&\frac 1{\alpha'} \bigg[ m(a_0^i {}^\dagger a_0^i +  
S_0^i {}^\dagger S_0^i) \nonumber \\  
&+& \left.\sum_{n=1}^\infty  
\sqrt{n^2+m^2} \left(\sum_{i=1}^8 (a_n^i {}^\dagger a_n^i +  
\tilde a_n^i {}^\dagger \tilde a_n^i)+ \sum_{a=1}^8 (S_n^a{}^\dagger  
S_n^a + \tilde S_n^a{}^\dagger \tilde S_n^a)\right)  
\right]  
\eea  
where we follow the conventions of Metsaev and Tseytlin \cite{metse}.   
We want to emphasize at this point that the  mass parameter  
is $p^+$-dependent:  
\be  
m=\mu (2\pi\alpha') p^+.  
\ee  
Explicitly performing the trace of the light-cone Hamiltonian, and  
implementing the level-matching   
constraint $\lambda(N_R-N_L)$ as discussed in the previous section  
leads to the light-cone partition function  
\bea  
\!\!\!\!\!\!\!z_{lc}(\frac\beta{p^+},\lambda)&=& \prod_{n,n'=1}^\infty  
\bigg[\exp   
\!\!\!\!\!\!\!\sum_{N,N'=0}^{\infty}\!\!(  
-\frac{\beta}{\alpha'p^+}N\sqrt{n^2+m^2} -\frac{\beta}{\alpha' p^+}  
N'\sqrt{n'^2+m^2}  
+ 2\pi i\lambda(Nn-N'n')) \bigg]^8\nonumber\\  
&&   
\bigg[\exp\sum_{N,N'=0,1}(-\frac{\beta}{\alpha' p^+} N\sqrt{n^2+m^2}   
-\frac{\beta}{\alpha'p^+}  
N'\sqrt{n'^2+m^2}  
+ 2\pi i\lambda(Nn-N'n')) \bigg]^8\nonumber\\  
&&\bigg[\exp\sum_{N=0}^\infty(-\frac{\beta}{\alpha' p^+} Nm)\bigg]^8  
\bigg[\exp\sum_{N=0,1}(-\frac{\beta}{\alpha' p^+} Nm)\bigg]^8\nonumber\\  
&=&\prod_{n=1}^\infty \left|  
\left(\frac{1+\exp(-\frac{\beta}{\alpha'p^+}\sqrt{n^2+m^2}  
+2\pi i\lambda n)}{1-\exp(-\frac{\beta}{\alpha'p^+}\sqrt{n^2+m^2}  
+2\pi i\lambda n)}\right)^8  
\right|^2 \left(\frac{1+\exp(-\frac{\beta}{ \alpha'p^+} m)}  
{1-\exp(-\frac{\beta}{\alpha' p^+} m)}\right)^8  
\nonumber\\  
&=& \prod_{n\in {\bf Z}}  
\left(\frac{1+\exp(-\frac{\beta}{\alpha'p^+}\sqrt{n^2+m^2}  
+2\pi i\lambda n)}{1-\exp(-\frac{\beta}{\alpha'p^+}\sqrt{n^2+m^2}  
+2\pi i\lambda n)}\right)^8~~~~,  
\label{zlc}  
\eea  
where we have taken into account that the usual normal ordering shift  
cancels between fermions and bosons (in each sector it equals the  
zero-point energy $m/2 + \sum_{n=1}^\infty \sqrt{n^2+m^2}$).   
  
The above expression is formally the answer for the partition   
function. However, it is very difficult to extract information   
directly in this form. We turn to the path integral approach  
where we can derive    
the modular properties of partition   
function. The latter will enable us   
to study the thermodynamic properties  
of the strings.   
  
%%%%%%%%%%%%%%%%%%%%%%%%%%%%%%%%%%%%%%%%%%%%%%%%%%%%%%%%%%%%%%  
\subsection{Scalars}  
%%%%%%%%%%%%%%%%%%%%%%%%%%%%%%%%%%%%%%%%%%%%%%%%%%%%%%%%%%%%%%%%%%  
  
To motivate some of the further manipulations needed to extract information  
from the partition function presented above   
we will also include the treatment for flat space.  
Another important point we would like to make is that by considering strings  
at nonzero temperature we explicitly break supersymmetry and  
therefore some of the thermodynamic properties of strings can be  
seen equally well by considering only the bosonic sector.  
Thus, we analyze first the bosonic  
contribution to the free energy, keeping in mind similarities with  
a system of closed bosonic strings. Let us briefly recall the determination  
of the Hagedorn temperature for the bosonic string in flat space.   
Each scalar degree of freedom  contributes the following   
factor to the the transverse  
partition function  
\bea  
z_{lc}^{(0,0)}(\frac\beta{p^+},\lambda)&=&  
\int_{-\infty}^\infty dp  
\exp(-\frac{\beta p^2}{2\alpha' p^+})  
\exp(\frac\beta{\alpha' p^+} \sum_{n=1}^\infty n)  
\prod_{n=1}^\infty \left|  
\frac{1}{1-\exp(-\frac\beta{\a'\,p^+} n + 2\pi i\lambda n)}\right|^2\nonumber\\  
&=&2\sqrt{\frac{\pi \a'\,p^+}{\beta}}\exp(-\frac{1}{12}  
\frac\beta{\a'\, p^+})  
\prod_{n=1}^\infty \left|   
\frac{1}{1-\exp(-\frac\beta{\a'\, p^+} n + 2\pi i\lambda n)}\right|^2.  
\eea  
A convenient way of rewriting the above expression is   
by means of introducing a complex variable  
\be  
q=\exp(2\pi i \tau)~~~~,~~~~ \tau=\lambda+ i \frac\beta{2\pi \alpha'p^+}.  
\ee  
In this variable the partition function takes the following simple   
form which highlights its modular properties   
\be   
z_{lc}^{(0,0)}(\tau)  
=\tau_2^{-1/2}\left((q \bar q)^{\frac{1}{24}}   
\left|\prod_{n=1}^\infty(1-q^n)\right|   
\right)^{-2}=\tau_2^{-1/2}|\eta(\tau)|^{-2}.  
\ee  
For $d-2$ transverse scalars, the one-string partition function  
reads  
\bea  
Z_1^{(0,0)}(\beta)&=&l_s \int_0^\infty dp^+ \int_0^1 d\lambda \exp(-\beta p^+)  
\left(  
\sqrt{\frac{2\pi\a'\, p^+}{\beta}}|\eta(\lambda + i\frac{\beta}{2\pi p^+})|^{-2}  
\right)^{d-2}  
\nonumber\\  
&=&\frac{\beta}{2\pi l_s}  
\int d\tau_1 \int_0^\infty d\tau_2 \exp(-\frac{\beta^2}{2\pi\a'\,\tau_2})  
2^{d-2} \tau_2^{-\frac{d-2}{2}+2}  
|\eta(\tau)|^{-2(d-2)}\label{scalar},  
\eea  
where we have made the change of variable   
$\frac{\beta}{2\pi\, \alpha'\, p^+}=\tau_2$.   
The UV    
asymptotic behavior, $\tau_2\rightarrow 0$,  
of the one-string partition function is uncovered by   
using the modular properties  
of the Dedekind  
$\eta(\tau)$-function. Under the $S$-modular  
transformation $\tau \rightarrow -1/\tau$  
\be  
\eta(-\frac{1}{\tau})=\sqrt{-i\tau}\eta(\tau)\label{s}.  
\ee  
Substituting (\ref{s}) into (\ref{scalar}) one finds that the integrand  
of the one-string partition function in the UV regime behaves as  
\be  
\tau_2^{-\frac{d+2}{2}} |\tau|^{d-2}  
\exp\left(4\pi\tau_2\frac{d-2}{24 |\tau|^2}   
- \frac{\beta^2}{2\pi \alpha'\, \tau_2}  
\right).  
\ee  
At temperatures higher than the  Hagedorn temperature  
\be  
\beta_H=\frac{1}{T_H}=\sqrt{\frac{\pi^2 (d-2)}{3}} l_s  
\ee  
the free energy diverges.  
  
To evaluate the contribution of the scalar modes of the IIB GS  
superstring in the plane wave background we choose to do a path integral  
calculation. This approach will pay off in the sense that the modular  
properties of the partition function are much more transparent  
when the partition function is expressed as a double product.  
Let us therefore begin with  
\be  
z_{lc}^{(0,0)}(\tau, m)=  
\int {\cal D}X \exp\bigg[-\int_T d^2 z \bar{X}(-\partial_z\partial_{\bar z}  
+m^2)X\bigg],  
\label{path}  
\ee  
where the worldsheet integral is taken over a torus: $z=\xi_1+\tau \xi_2$,  
and the modular parameter is denoted as usual by $\tau$.  
Substituting the Fourier decomposition of the doubly periodic  
function  
\be  
X(\xi_1, \xi_2)=\sum_{n1,n2\in {\bf Z}}X_{n1,n2}  
\exp[2\pi i(n_1\xi_1+n_2\xi_2)]  
\ee  
in the path integral and using  
\bea  
d^2 z&=& d\xi_1 d\xi_2 \tau_2, \\  
\partial_z\partial_{\bar z}&=&\frac{1}{4\tau_2{}^2}\left(|\tau|^2\partial_1{}^2  
-2\tau_1\partial_1\partial_2+\partial_2{}^2\right),  
\eea  
we can explicitly perform the Gaussian integrals over $X_{n1,n2}$  
in (\ref{path})  
\be  
z_{lc}^{(0,0)}(\tau, \mu) = \left[\prod_{n_1,n_2\in {\bf Z}}  
\tau_2 \left({(\frac{2\pi}{4\tau_2})^2 |n_1 \tau - n_2|^2 +m^2  
}\right)\right]^{-(d-2)}.\label{path2}  
\ee  
The double product form of the partition function makes manifest   
its modular properties. For example, recalling that $m$   
depends explicitly on $p^+=\frac\beta{2\pi\, \alpha' \, \tau_2}$ and   
substituting $m=2\pi\alpha'\, \mu p^+$ into (\ref{path2}) one derives:  
\be  
z_{lc}^{(0,0)}(-1/\tau,\mu/|\tau|)) = z_{lc}^{(0,0)}(\tau, \mu). \label{sf1}  
\ee  
For cases considered in the literature \cite{bgg,takayanagi},  
$m$ was uncorrelated with the   
torus modular parameter,  
and (\ref{sf1}) is then replaced by $z_{lc}(-1/\tau, m |\tau|) = z_{lc}(\tau, m)$  
\footnote{  
After trivial relabellings in the double product,  
$$z_{lc}^{(0,0)}(-1/\tau,\mu/|\tau|)= \left[\prod_{n_1,n_2\in {\bf Z}}  
\frac{\tau_2}{|\tau|^2}  
\left({(\frac{2\pi|\tau|^2}{4\tau_2})^2  
\frac{|-n_1  - n_2\tau|^2}{ |\tau|^2 }  
+\left(\frac{\beta\mu|\tau|^2}{|\tau|\tau_2}  
\right)^2  
}\right)\right]^{-(d-2)}=z_{lc}^{(0,0)}(\tau, \mu)$$  
whereas for constant $m$ we have  
$$z_{lc}^{(0,0)}(-1/\tau, m|\tau|)=\left[\prod_{n_1,n_2\in {\bf Z}}  
\frac{\tau_2}{|\tau|^2}  
\left({(\frac{2\pi|\tau|^2}{4\tau_2})^2  
\frac{|-n_1  - n_2\tau|^2}{ |\tau|^2} +m^2 |\tau|^2  
}\right)\right]^{-(d-2)}  
=z_{lc}^{(0,0)}(\tau, m)$$  
}.  
  
Following \cite{is}, one of the infinite products in (\ref{path2})  
can be performed and the result is  
\bea  
z_{lc}^{(0,0)}(\tau,m)&=&\exp\left[  
-2\pi (d-2)\tau_2\left(m/2+\sum_{n=1}^\infty \sqrt{n^2+m^2}\right)\right]  
\nonumber\\  
&&\bigg[\prod_{n\in {\bf Z}}\left(1-\exp[2\pi (-\tau_2 \sqrt{n^2+m^2}+i\tau_1 n)]  
\right)\bigg]^{-(d-2)}.\label{z}  
\eea  
This expression is in fact a natural generalization of  
the Dedekind $\eta$ function,  
where the first exponent substitutes the factor $q^{1/24}$ of the  
$\eta$-function. We use $\zeta$-function regularization  
to define the  Casimir energy of a system of massive oscillators  
\bea  
\gamma_0(m)&=& \frac{m}2+\sum_{n=1}^\infty  
\sqrt{n^2+m^2} \nonumber \\  
&=& \frac{m}{2}+\left[-\frac{1}{12}   
+ \frac 12 m - \frac 12 m^2 \ln(4\pi e^{-\gamma})  
+ \sum_{n=2}^\infty (-1)^n  
\frac{\Gamma(n-\frac 12)}{n!\Gamma(-\frac 12)}  
\zeta(2n-1) m^{2n}\right]\label{gamma0},  
\eea  
where $\gamma$ is the Euler constant.   
Note that the zero-th order term in the $\gamma_0(m)$ mass expansion is  
the same as for a massless scalar $\sum_{n=1}^\infty n = \zeta(-1)=-1/{12}$.  
The terms of order greater and equal to 2 can be derived  
as follows:  
\bea  
\sum_{n=1}^\infty \sqrt{n^2+m^2}&=&\sum_{n=1}^\infty\frac{1}{\Gamma(-1/2)}  
\int_0^\infty dt \exp\left(-t(n^2+m^2)\right) t^{-3/2} \nonumber\\  
&=&\frac{1}{2\Gamma(-1/2)}  
\int_0^\infty dt \left[\Theta_3(0, \frac{it}{\pi}) -1\right] \exp(-t m^2)  
t^{-3/2},  
\eea  
and after Taylor expanding the exponential, use further that  
\be  
\int_0^\infty dx x^{s-1}\left[\Theta_3(0, i x^2) -1\right] =  
\pi^{-s/2}\Gamma(\frac{s}{2})\zeta(s),~~~~~~~s\ge 2.  
\ee  
The expression obtained in (\ref{z}) for the partition function also represents  
the generalization to the complex plane of the modular forms  
$f_1^{(m)} (e^{-2\pi\tau_2})$ introduced by \cite{bgg}. The feature   
that makes this generalization special is that the appropriate   
modular form for massive transverse scalars (as opposed to massless   
transverse scalars) is not a   
holomorphic function of the complex torus parameter $\tau$.   
  
The UV behavior of the scalar degrees of freedom of the closed  
one-string partition function  
\be  
Z_1(\beta,\mu)={\beta\over 2\pi l_s} \int_0^\infty   
\frac{d\tau_2}{\tau_2{}^2}\int d\tau_1  
\exp(-\frac{\beta^2}{2\pi\alpha'\tau_2})  
z_{lc}^{(0,0)}(\tau,\frac{\mu\,\beta}{\tau_2})\label{Z_1}.  
\ee  
is now determined   
with the help of the S-modular transformation (\ref{sf1}).   
The integrand of (\ref{Z_1}) behaves as  
\bea  
\label{1}  
\frac {1}{\tau_2^2}e^{-\frac{\beta^2}{2\pi\alpha'\,\tau_2}}  
z_{lc}^{(0,0)}(\tau, \mu){\stackrel{\tau_2\rightarrow 0}{\longrightarrow}}  
%\nonumber\\  
\frac {1}{\tau_2^2}e^{-\frac{\beta^2}{2\pi\alpha' \tau_2}}  
\left(1-e^{-\frac{\mu\beta}{|\tau|}}\right)^{-\frac{d-2}2}  
e^{-2\pi(d-2)\frac{\tau_2}{|\tau|^2}\gamma_{0}(\mu\,\beta)}.  
\eea  
Ignoring for   
the moment the fermionic degrees  
of freedom and judging only by (\ref{Z_1}), the Hagedorn phase transition  
occurs again at the point where $Z_1(\beta,\mu)$ becomes divergent  
\be  
-\frac{\beta_H^2}{2\pi\, \alpha'}-2\pi(d-2)\gamma_0(\mu\, \beta_H)=0.  
\ee  
  
Note that since the mass parameter $m$  
is temperature dependent $m=\mu\, \beta/\tau_2$, the finite  
temperature behavior of strings in the pp-wave background is  
fundamentally different than that of strings in flat space.  
We will come back to address this issue at length in Section 4.

%%%%%%%%%%%%%%%%%%%%%%%%%%%%%%%%%%%%%%%%%%%%%%%%%%%%%%%%%%%%%  
\subsection{Fermions}  
%%%%%%%%%%%%%%%%%%%%%%%%%%%%%%%%%%%%%%%%%%%%%%%%%%%%%%%%%%%%%%  
  
Let us briefly recall the thermal partition function of closed superstrings   
in flat space. The contribution of the fermionic (physical) degrees of freedom  
to the light-cone partition function     
\be  
z_{lc}^{(1/2,0)}(\tau)=  
\left(|q|^{\frac{1}{12}}\prod_{n=1}^\infty  
\bigg| 1+q^n\bigg|^2 \right)^{d-2}  
\ee  
can be rewritten using Jacobi's triple product formula in terms of the  
$\Theta$-functions:  
\bea  
\prod_{n=0}^\infty q^{\frac 1{24}}(1+q^n)&=&\eta(\tau)^{-\frac {1}2}  
\left[\prod_{n=0}^\infty (1+q^{n+1})^2 (1-q^n)\right]^{\frac 12}=  
\left( \Theta_2(0,\tau) \eta(\tau)^{-1} \right)^{1/2}  
\eea  
For type II superstrings ($d-2=8$) the one string   
partition function is  
\be  
Z_1^{(1/2,0)}(\beta)  
={\beta \over 2\pi l_s} \int_0^\infty \frac{d\tau_2}{\tau_2{}^2}  
\int d\tau_1  
\exp(-\frac{\beta^2}{2\pi\alpha'\tau_2}) 2^{-16}\left|\Theta_2(0,\tau)   
\eta(\tau)^{-3}\right|^8.  
\ee  
Using  some $\Theta$-function algebra:   
$\Theta_2 (0,\tau)\Theta_3 (0,\tau)   
\Theta_4 (0,\tau) = \eta(\tau)^3$ and $\Theta_3(0,\tau) \Theta_4(0,\tau)=  
\Theta_4(0,2\tau)$  
the one-string partition function can be cast in a more concise form   
\be  
Z_1^{(1/2,0)}(\beta)  
=\beta \int_0^\infty \frac{d\tau_2}{\tau_2{}^2}\int d\tau_1  
\exp(-\frac{\beta^2}{2\pi\alpha'\tau_2})2^{-16} |\Theta_4(0,2\tau)|^{-16}  
\label{Zz1}.  
\ee  
Next, by  invoking the modular properties of $\Theta_4$  
\be  
\Theta_4(0,\tau)=(-i\tau)^{-\frac 12}\Theta_2(0,-\frac 1\tau)  
\ee  
one obtains the ultraviolet behavior of the integrand in (\ref{Zz1}):  
\be  
\tau_2^2 e^{-\frac{\beta^2}{2\pi\alpha'\tau_2}} e^{\frac{2\pi}{\tau_2}}.  
\ee  
Since near $\tau_2\rightarrow 0$ the integrand is convergent,  
the type II superstrings in flat space   
background can undergo a phase transition at $T_H=1/(2\pi l_s )$.

In the plane wave background the transverse fermions are massive   
and this  
prevents us from using the triple product formula or the   
$\Theta$-function technology. However, the  
fermionic partition function is still a modular form, and we can still  
perform an $S$-transformation in order to read-off its asymptotics  
for  $\tau_2\rightarrow 0$.

The light-cone gauge-fixed action of the   
fermionic degrees of freedom for  type IIB GS superstring in the  
plane wave background takes the form  
\footnote{  
A few manipulations are in order to bring the fermionic action in this  
form. One first redefines the Weyl complex space-time fermions   
of \cite{metse} by  
conveniently absorbing a factor of $\sqrt{p^+}$:   
$\theta^\alpha=\frac{S_1^\alpha+i S_2^\alpha}{\sqrt {2p^+}},  
\alpha = 1..8$   
and after evaluating the matrix $\Pi=\gamma^1\gamma^2\gamma^3\gamma^4$  
present in the fermionic mass term $m p^+\bar\theta\Pi\theta$, a trivial  
re-shuffling of the variables $S^\alpha$ leads to the action presented  
here.   
}  
\be  
{\cal S}=\int d^2 z\sum_{a=1}^8\left(  
i S^{1,a}\partial_z S^{1,a} + iS^{2,a}\partial_{\bar z}  
S^{2,a} + 2m S^{1,a}S^{2,a}\right).  
\ee  
  
Since we are interested in evaluating the thermal  
partition function, we impose periodicity  
in the $\sigma$-direction and antiperiodicity in the world-sheet   
time direction. Therefore we make the identifications  
$S^a(\xi_1+1,\xi_2)=S^{a}(\xi_1,\xi_2)$, $S^a(\xi_1,\xi_2+1)=  
-S^a(\xi_1,\xi_2)$,  
$z=\xi_1+\tau\xi_2$, where as explained before  
$\tau=\lambda+i\frac{\beta}{2\pi\, \alpha ' p^+}$. In the new coordinates  
the action becomes  
\be  
{\cal S}=\int d \xi_1 d\xi_2 \left(iS^{1,a}(-\bar\tau \partial_1 +\partial_2)  
S^{1,a} + iS^{2,a}(\tau \partial_1 -\partial_2)  
S^{2,a} + 2m S^{1,a}S^{2,a}\right).  
\ee  
Substituting the appropriate Fourier decomposition of the fermionic  
degrees of freedom   
\be  
S^a(\xi_1,\xi_2)=\sum_{n_1, n_2 \in{\bf Z}}S^a_{n_1,n_2}  
e^{i 2\pi n_1\xi_1} e^{i \pi  (2 n_2 + 1)\xi_2}  
\ee  
into the the path integral  
expression of the partition function we obtain  
\bea  
\!\!\!\!\!\!\!\!\!\!\!\!  
z_{lc}^{(0,1/2)}  
(\tau, m)&=&\int {\cal D}S^1 {\cal D}S^2 e^{-{\cal S}[S^1, S^2]}  
\nonumber\\  
&=&\left[\prod_{n_1,n_2\in {\bf Z}}  
\tau_2 \left({(\frac{2\pi}{4\tau_2})^2 \left|n_1 \tau + \frac{2 n_2+1}2  
\right|^2 +m^2  
}\right)\right]^8\nonumber\\  
&=& \exp\left[  
16\pi \tau_2\left(\frac m2+\sum_{n=1}^\infty \sqrt{n^2+m^2}\right)  
\right]\left\{  
\prod_{n\in{\bf Z}}\left(1+\exp[2\pi (-\tau_2 \sqrt{n^2+m^2}  
+i\tau_1 n)]\right)\right\}^8\nonumber\\  
\label{path3}  
\eea  
where in the last step we used a result of \cite{is}.  
  
Under the modular $S$-transformation, a function antiperiodic in the  
$\xi_2$ direction becomes antiperiodic in the $\xi_1$ direction.  
The partition function for fermions obeying this new boundary conditions  
is  
\bea  
\!\!\!\!\!\!\!\!\!\!\!\!  
z_{lc}^{(1/2,0)}(\tau,m)&=&\left[\prod_{n_1,n_2\in {\bf Z}}  
\tau_2 \left({(\frac{2\pi}{4\tau_2})^2 \left|\frac{2 n_1+1}2 \tau  
+ n_2\right|^2 +m^2  
}\right)\right]^8\\  
&=&\exp\left[  
16\pi\tau_2\sum_{n=1/2}^\infty \sqrt{n^2+m^2}\right]\left\{  
\prod_{n\in{\bf Z}+1/2} \left(1-\exp[2\pi (-\tau_2 \sqrt{n^2+m^2}  
+i\tau_1 n)]\right)\right\}^8. \nonumber   
\label{path4}  
\eea  
The Casimir energy for half-integer modes can be   
computed using $\zeta$-function regularization   
\bea  
\gamma_{1/2}(m)&=&\sum_{n=1/2}^\infty  
\sqrt{n^2+m^2}=\frac 12\sum_{n=1}^\infty \sqrt{(2n+1)^2+4m^2}=  
\frac 12 \sum_{n=1}^\infty\sqrt{n^2+4m^2}-\sum_{n=1}^\infty\sqrt{n^2+m^2}  
\nonumber\\  
&=&{1\over 2}\big[\gamma_0(2m)-m\big]-\big[\gamma_0(m)-{m\over 2}\big]   
\nonumber \\  
&=&\frac{1}{24} - \frac 12 m^2 \ln(\pi e^{-\gamma})  
+ \sum_{n=2}^\infty (-1)^n \left(2^{2n-1}-1\right)  
\frac{\Gamma(n-\frac 12)}{n!\Gamma(-\frac 12)}  
\zeta(2n-1) m^{2n},\label{gamma12}  
\eea  
where in the first line we rewrite the sum over odd integers as the  
the sum over all minus even integers;    
in the second line we use the $\zeta$-function  
regularization introduced in (\ref{gamma0}).   
As we expected,  
the modular $S$-transformation relates the two partition  
functions  
\be  
z_{lc}^{(1/2,0)}(1/\tau,\mu/|\tau|)=z_{lc}^{(0,1/2)}(\tau,\mu).  
\ee  
Once again, recall that $m$ is not a constant: $m=\mu\,\beta/\tau_2$.  
We can now easily extract the high-energy behavior of the fermionic  
light-cone partition function:  
\be  
\label{2}  
z_{lc}^{(0,1/2)}(\tau, \mu){\stackrel{\tau_2\rightarrow 0}{\longrightarrow}}  
e^{16\pi\frac{\tau_2}{|\tau|^2}\gamma_{1/2}(\mu\,\beta)}.  
\ee  
The two modular forms we have introduced in (\ref{path3}) and (\ref{path4})  
are also generalization to the complex plane of the modular  
forms $f_2^{(m)}\left(e^{-2\pi\tau_2}\right)$ and respectively  
$f_4^{(m)}\left(e^{-2\pi\tau_2}\right)$ of \cite{bgg} with   
the distinguishing property that they are   
not holomorphic functions of $\tau$.   
  
%%%%%%%%%%%%%%%%%%%%%%%%%%%%%%%%%%%%%%%%%%%%%%%%%%%%%%%%%%%%%  
\section{Hagedorn temperature of IIB strings in the pp-wave background}  
%%%%%%%%%%%%%%%%%%%%%%%%%%%%%%%%%%%%%%%%%%%%%%%%%%%%%%%%%%%%%%%%%%%  
We have by now computed all the ingredients that go into   
the partition function. In a sense we can now simply return  
to the light-cone string partition function already given   
by (\ref{zlc}). However, after the detour of the path integral  
computation we have gained knowledge of the modular properties of (\ref{zlc}),  
and thus paved the road to   
 find out the high energy behavior of the partition function.   
The one-string partition function is   
\be  
Z_1(\beta,\mu)={\beta\over 2\pi \, l_s}  
\int_0^\infty \frac{d\tau_2}{\tau_2{}^2}\int d\tau_1  
\exp(-\frac{\beta^2}{2\pi\,\alpha'\,\tau_2})  
z_{lc}^{(0,0)}(\tau,\frac{\mu\,\beta}{\tau_2})\,  
z_{lc}^{(0,1/2)}(\tau,\frac{\mu\,\beta}{\tau_2})  
\label{Z_1f}.  
\ee  
In terms of the one-string partition function the   
free energy is   
\bea  
F(\beta) &=&-{1\over 2 \beta}\sum\limits_{r=1}^\infty  
\big[1-(-1)^r\big]{Z_1(\beta\, r)\over r} \nonumber \\  
&=& -{1\over 4\pi\, l_s}\int_0^\infty \frac{d\tau_2}{\tau_2{}^2}  
\int d\tau_1 \sum\limits_{r=1}^\infty   
\big[1-(-1)^r\big]  
e^{-\frac{\beta^2\, r^2}{2\pi\,\alpha'\,\tau_2}}  
z_{lc}^{(0,0)}(\tau,\frac{\mu\,\beta\, r}{\tau_2})\,  
z_{lc}^{(0,1/2)}(\tau,\frac{\mu\,\beta\,r}{\tau_2})\label{free}.  
\eea  
In flat space where the only dependence on the number of   
strings $r$  comes from   
$\exp(-\beta^2 r^2/(2\pi\alpha'\tau_2)$,   
one can explicitly perform the summation   
over $r$ and write the free energy as   
a combination of theta functions \cite{a2}. Given the more involved $r$  
dependence in (\ref{free}) we have not been able   
to find such a closed form of the free energy. However, it can still be   
argued that the main contribution arises from the   
one-string term. In this case the relevant integrand behaves as   
(\ref{1}, \ref{2}):  
\bea  
&&\frac{1}{\tau_2^2}  
e^{-\frac{\beta^2}{2\pi\,\alpha' \tau_2}}   
z_{lc}^{(0,0)}(\tau,\frac{\mu\,\beta}  
{\tau_2})z_{lc}^{(0,1/2)}(\tau,\frac{\mu\,\beta}{\tau_2})  
{\stackrel{\tau_2\rightarrow 0}{\longrightarrow}}\nonumber\\  
 &&\frac{1}{\tau_2^2}e^{-\frac{\beta^2}{2\pi\alpha'\tau_2}}  
\left(1-e^{-\frac{\mu\,\beta}{|\tau|}}\right)^{-4}  
e^{-16\pi\frac{\tau_2}{|\tau|^2}\gamma_{0}(\mu\,\beta)}  
e^{16\pi\frac{\tau_2}{|\tau|^2}\gamma_{1/2}(\mu\,\beta)}\label{mod},  
\eea  
where we have used the S-modular transformation to arrive to the   
second line. Considering the region where the above expression   
diverges we find that   
the Hagedorn temperature for the GS IIB superstring in the plane   
wave background  
is determined by solving:  
\be  
\label{hagedorn}  
-\frac{\beta_H^2}{2\pi\,\alpha'}-16 \pi\gamma_0(\mu\,\beta_H)  
+16 \pi \gamma_{1/2}  
(\mu\,\beta_H)=0,  
\ee  
and it depends parametrically on $\mu$. Taking $\m=0$ in the above   
equation and making use of (\ref{gamma0}) and (\ref{gamma12}) we find that  
$\beta_H=2\pi\, l_s$ which is nothing but the value corresponding   
to flat space. To further explore the behavior of the   
Hagedorn temperature we consider the limit of small   
$\m$ in units of the string scale $l_s$, that is,  $\m\,l_s \ll 1$. Expanding  
(\ref{hagedorn}) up to terms linear in $\m$ and using the   
expressions (\ref{gamma0}, \ref{gamma12}) we find   
\be  
T_H\approx{1\over 2\pi l_s} + 4 \m, \qquad \rm{for} \quad    \m\, l_s \ll 1,  
\ee  
that is, the Hagedorn temperature is higher than its value   
for strings in flat space.   
  
This result is very interesting and naturally leads  us into the   
question  of the behavior in the opposite limit, $\m l_s \to  
\infty$. This limit is the more important due   
to the results of \cite{gauge} where it has been established   
that in the context of the BMN limit it corresponds to the   
weakly coupled field theory.\footnote{However, as mentioned   
in the introduction,   
we should be extremely careful conjecturing the relation of   
our results to the BMN limit since we simply do not work on that   
limit.}   
To better answer the question of the Hagedorn temperature   
for large $m$ we find it convenient to use an alternative expression  
for the Casimir energies (see the appendix for a derivation):  
\bea  
\gamma_0(m)=\frac m2 + \sum_{n=1}^\infty\sqrt{n^2+m^2}=  
\frac{m}2+\left[  
-\frac{m}2 -  
\frac{-1+\gamma}{4}m^2 -  
\frac{m}{\pi}\sum_{n=1}^\infty \frac{1}{n}K_{-1}(2\pi n m)\right].\label{gk}  
\eea   
For the fermions we will simply use the identity   
described in (\ref{gamma12}) to relate their  Casimir energy to that of  
bosons. In equation (\ref{hagedorn}) we have that $\beta_H$ is proportional to   
the difference of the Casimir energies:  
\bea  
{\beta_H^2\over 2\pi\, \alpha'}&=&16\pi   
\big[\gamma_{1/2}(\m\, \b_H) -\gamma_{0}(\m\, \b_H)\big]  
= {1\over 2}\gamma_0(2 \, \m\, \b_H )-2\gamma_0(\m\, \b_H) \nonumber \\  
&=&{\m \, \b_H\over \pi}\sum\limits_{n=1}^\infty {1\over n}  
\bigg[2 K_{-1}(2\pi\, n\, \m\b_H)-K_{-1}(4 \pi\, n\, \m\b_H)\bigg].  
\eea  
As $\mu\rightarrow\infty$, the difference between the two Casimir energies  
approaches zero since all the Bessel functions  
in the infinite sum (\ref{gk}) vanish in the limit, and therefore   
the  Hagedorn temperature is pushed toward infinity.    
Recalling that the Hagedorn temperature   
can be interpreted as the temperature where   
a winding state becomes tachyonic \cite{pat}\footnote{From the   
path integral approach we obtain the free energy naturally expressed   
as an integral over the entire strip $(-1/2\le \tau_1 \le 1/2,\,  
0\le \tau_2 < \infty)$. Using the fact that the integrand enjoys   
modular properties, the free energy can be rewritten as an integral   
over the fundamental domain only. The net effect of this manipulation  
is to add a trace over ``winding'' modes in the partition function.   
These modes can be interpreted as states of the string wrapping the   
temporal direction. },  
we can rephrase the above result  as  the disappearance of the   
tachyon associated with the Hagedorn temperature.  
In this form the result is similar to recent investigations   
of type 0B \cite{takayanagi,0b}, where the tachyon has been   
found to disappear in the $\mu\rightarrow\infty$ limit.   
  
Having established the existence and some of the properties of the   
Hagedorn temperature we turn to the question of its nature. Namely, we  
would like to find out whether the Hagedorn temperature signals a   
maximal temperature or it simply points to the possibility of a phase   
transition. To answer this question we should consider the behavior of   
the free energy and the specific heat as we approach the   
Hagedorn temperature from below. Working in the one-string approximation   
and assuming that the main contribution to the free energy (\ref{free})  
comes form the UV region and using (\ref{mod}), we find  
that, for fixed $\mu$, the free energy diverges near the Hagedorn temperature  
\be  
F\sim -{1\over 2\pi l_s}\int\limits_0^1 d\tau_2  
\frac{1}{\tau_2^2} \exp \left(-{1\over \tau_2}\bigg[{\beta^2\over 2\pi \alpha'}  
+16\pi(\gamma_0(\m\, \beta)-\gamma_{1/2}(\m\,\beta))\bigg] \right).  
\label{nearhag}  
\ee  
Evaluating the integral we can estimate the behavior of the free energy near   
the Hagedorn temperature  
\be  
F(T)\sim C(\b_H)^{-1} \frac{T_H}{T_H-T}  
+ \mbox{analytic in}\,\, (\b-\b_H).\label{approx}  
\ee  
where $C(\b_H)=\b_H^2/(\pi\, \a')   
+16\pi\b_H\mu [\gamma'_0(\m\b_H)-\gamma'_{1/2}(\m\b_H)]$.  
Let us now turn to the specific heat which we write in terms of the   
free energy as  
\be  
c_V=\beta^2{\partial^2 \over \partial\beta^2}\ln Z=  
-\b^2{\partial^2 \over \partial\beta^2}\b F  
=-\b^2\bigg[2{\partial \over \partial\beta}F  
+\b{\partial^2 \over \partial\beta^2}F\bigg].  
\ee  
Using (\ref{approx}) one derives that the specific heat also  
blows up near the Hagedorn temperature. Therefore for IIB strings   
in RR plane wave background the Hagedorn temperature is limiting,  
and, at least from the point of view of a canonical ensemble analysis,  
there is no phase transition which the system undergoes at  
$T_H$. Curiously, in flat space only open strings seem to have a  
limiting Hagedorn temperature. At least to this level of scrutiny,  
IIB strings in the RR plane wave background behave as open strings.   
\vskip 0.5cm  

{\bf Note added:}

\vskip 0.5cm

A better estimation of the free energy behavior (finite or divergent) at the Hagedorn temperature
can be done by evaluating the $\tau_1$ integral, as pointed out by Brower et al. 
\cite{brower}. Using the saddle point approximation, the integral over $\tau_1$ of (\ref{mod})
yields
\bea
F&\sim&-\frac{1}{2\pi l_s} \int d\tau_2  ~\frac{1}{\tau_2^2}  
e^{-\frac{\beta^2}{2\pi\,\alpha' \tau_2}}\int d\tau_1~
e^{  \frac{\beta_H^2}{2\pi\,\alpha' \tau_2}}\e^{-\frac{\beta_H^2\tau_1^2}{2\pi\alpha'\tau_2^3}}
\nonumber\\
&\sim&-\frac{1}{\beta_H}\int d\tau_2  ~\tau_2^{-1/2}e^{\frac{\beta_H^2-\beta^2}
{2\pi\,\alpha' \tau_2}}\sim -c(\beta_H) \sqrt{\beta-\beta_H}+\mbox{regular}
\eea
which should be compared with the rougher  approximation used in (\ref{nearhag}), where the 
integral over $\tau_1$ was omitted.

Thus the free energy is, in the end, finite, and a phase transition may occur when the system reaches the
Hagedorn temperature.
Note however that the specific heat is infinite at $T_H$, and negative.

 \vskip 3cm
\begin{center}  
{\large Acknowledgments}  
\end{center}  
We are grateful to G. D'Appollonio, R. Emparan, A. Hashimoto,  
J. Maldacena, N. Quiroz, J. Sonnenschein, C. Thorn and A. Tseytlin.    
We are grateful to B. Greene, K. Schalm and G. Shiu for comments on a  
previous version of the manuscript and for informing us of their related work.  
We are especially thankful to   
Igor Klebanov for a very illuminating discussion. The work of D.V. is   
supported by DOE grant DE-FG02-91ER40671.

\appendix{}  
  
For completeness we include a brief derivation of (\ref{gk}).  
\bea  
&&\sum_{n=1}^\infty\sqrt{n^2+m^2}=\frac12\left  
(\sum_{n\in{\bf Z}}\sqrt{n^2+m^2}-m\right)=\frac 12 [F(0)-m]\label{gka}  
\\  
&&F(a)=\sum_{n\in{\bf Z}}\sqrt{(n+a)^2+m^2}  
\eea  
Using the Poisson summation formula we can write  
\be  
F(a)=\sum_{k\in{\bf Z}}\exp(2\pi i k a)\int_{-\infty}^\infty  
dy \exp(-2\pi i k y)\sqrt{y^2+m^2}\label{f1}.  
\ee  
Next substitute  
\be  
\sqrt{y^2+m^2}=\frac{1}{\Gamma(-1/2)}  
\int_0^\infty\, dt\; t^{-3/2} \exp[-t(y^2+m^2)]  
\ee    
into (\ref{f1}), perform the Gaussian integral over $y$ and   
arrive at  
\be  
F(a)=\frac{\sqrt \pi}{\Gamma(-1/2)}\sum_{k\in {\bf Z}}  
\exp(2\pi i k a)\int_0^\infty dt \;t^{-2}\exp(-tm^2-\frac{k^2\pi^2}t).  
\label{f2}  
\ee  
The term $k=0$ in the sum leads to a divergent expression, proportional to  
$\Gamma(-1)/\Gamma(-1/2)$ which is further regulated by keeping only the   
finite part of $\Gamma(-1+\epsilon)$, with $\epsilon \rightarrow 0$.  
  
After one more change of variable $v=t m/(\pi |k|)$,   
the other terms in (\ref{f2}) can be expressed  
in terms of modified Bessel functions, and we obtain  
\be  
F(a)=-\frac{-1+\gamma}2 m^2 -\sum_{k=1}^\infty \frac{2m\cos(2\pi ka)}{\pi k}  
K_{-1}(2\pi km).  
\ee  
Substituting this expression into (\ref{gka}) completes the derivation of   
(\ref{gk}).

%%%%%%%%%%%%%%%%%%%%%%%%%%%%%%%%%%%%%%%%%%%%%%%%%%%%%%%%%%%%%%%  

\end{document}